\documentclass[prb,twocolumn,floatfix]{revtex4}
\usepackage{graphicx}

\def\lnod{La$_2$NiO$_{4+\delta}$}
\def\lsno{La$_{2-x}$Sr$_x$NiO$_4$}
\def\lsco{La$_{2-x}$Sr$_x$CuO$_4$}

\begin{document}

\title{Mid-Infrared Conductivity from Mid-Gap States Associated with
Charge Stripes}
\author{C. C. Homes}
\author{J. M. Tranquada}\email{jtran@bnl.gov}
\author{Q. Li}
\author{A. R. Moodenbaugh}
\affiliation{Brookhaven National Laboratory, Upton, NY  11973-5000} 
\author{D. J.~Buttrey}
\affiliation{Department of Chemical Engineering, University of Delaware,
Newark, Delaware 19716}
\date{March 12, 2003}
\begin{abstract} The optical conductivity of \lsno\ has been interpreted
in various ways, but so far the proposed interpretations have neglected
the fact that the holes doped into the NiO$_2$ planes order in diagonal
stripes, as established by neutron and X-ray scattering.  Here we present
a study of optical conductivity in La$_2$NiO$_{4+\delta}$ with
$\delta=\frac2{15}$, a material in which the charge stripes order
three-dimensionally.  We show that the conductivity can be decomposed
into two components, a mid-infrared peak that we attribute to transitions
from the filled valence band into empty mid-gap states associated with
the stripes, and a Drude peak that appears at higher temperatures as
carriers are thermally excited into the mid-gap states.  The shift of
the mid-IR peak to lower energy with increasing temperature is explained
in terms of the Franck-Condon effect.  The relevance of these results to
understanding the optical conductivity in the cuprates is discussed.
\end{abstract}
\pacs{PACS: 71.27.+a, 75.40.Cx, 75.50.Ee, 71.45.Lr}
\maketitle

\section{Introduction}

Studies of optical conductivity can provide valuable information about
electronic correlations in solids; however, the proper explanation for
the frequency dependence of the conductivity is not always uniquely
obvious.  One system that has received considerable attention is
La$_{2-x}$Sr$_x$NiO$_4$, a compound that is
isostructural with superconducting 
cuprates.\cite{bi90,ido91,bi93,cran93,eagl95,calv96,kats96,baum98,tsut99,%
kats99,pash00,jung01} 
The substitution of Sr$^{2+}$ for La$^{3+}$ dopes holes into the NiO$_2$
plane.  When the light polarization is  parallel to the NiO$_2$ planes,
the  electronic conductivity exhibits a broad peak at $\sim
0.6$~eV,\cite{bi90} whose intensity grows with the hole
concentration.\cite{ido91}   Katsufuji {\it et  al.}\cite{kats96} showed
for La$_{1.67}$Sr$_{0.33}$NiO$_4$ that spectacular changes occur on the
scale of 1 eV as the temperature is changed from 10 K up to 480~K, the
latter being twice the temperature at which the charge is know to
order.\cite{lee97} The mid-infrared (MIR) conductivity peak has been
variously interpreted in terms of
polarons,\cite{bi93,eagl95,calv96,kats96,baum98,jung01} $d$--$d$
excitations,\cite{tsut99,jung01} a doped-semiconductor,\cite{cran93} and 
conventional charge-density-wave (CDW) ordering\cite{kats96}; however,
none of these interpretations is particularly convincing.  

When the first infrared reflectivity studies were
performed,\cite{bi90,ido91} the nature of the electronic correlations in
\lsno\ was a mystery.  Today, however, it is firmly established by
neutron and X-ray diffraction studies that the holes doped into the 
NiO$_2$ planes tend to order in charge  stripes, separating
antiferromagnetic domains.\cite{tran94a,tran98b,yosh00}  Recently it has
been demonstrated that the charge stripes survive above the melting
transition.\cite{lee02} It follows that the optical conductivity must
derive from the electronic states associated with charge stripes. 

To provide context for our interpretation, we present an infrared
reflectivity study of a nickelate system that exhibits long-range,
three-dimensional (3D) stripe order\cite{tran95b,woch98}: \lnod\ with
$\delta=\frac2{15}$.  The 3D ordering of the oxygen interstitials creates
a modulated potential that can couple to the stripes.\cite{woch98} 
Based on measurements of the in-plane resistance, we argue that the
stripe and interstitial orderings appear simultaneously at 317~K, an
unusually high temperature for stripe order in the nickelates.

We analyze the optical conductivity
in terms of two components: a damped-harmonic oscillator (corresponding
to the MIR  peak) whose resonance frequency shifts with temperature, and
a Drude component.   We attribute the MIR peak to transitions from filled
valence states to empty  mid-gap states associated with the charge
stripes.\cite{zaan89,poil89,schu90,zaan96b,tche00,gran02,lore02,heeg88} 
The temperature-dependent energy shift of the MIR peak is due to the
Franck-Condon effect associated with lattice relaxation about the charge
stripes; a greater excitation energy is required in the ordered state
than in the melted state. The Drude  peak appears at higher temperatures
as the MIR shift is reduced and electrons  are thermally excited into the
mid-gap states.  We argue that the excitation of  carriers corresponds to
local fluctuations of the charge stripes, so that dc conductivity
competes with stripe order in this stripe insulator.  Such competition
would be less significant in a system with partially filled mid-gap 
states, as should be the case in certain cuprates with metallic stripes. 
\cite{noda99,ichi00,taji99,dumm02} 

The rest of this paper is organized as follows.  A description of the
experimental procedures in the next section is followed by a discussion
of the nature of stripe ordering in \lnod.  In
Sec.~\ref{sec:expect} we describe our expectations for the
optical conductivity associated with stripes based on available
theoretical calculations.  The experimental results for the optical
conductivity are presented and analyzed in
Sec.~\ref{sec:results}.  We conclude with a summary of our
results and interpretation, and a discussion of the relevance of this
picture to the interpretation of optical conductivity in cuprate
superconductors.

\section{Experimental Procedures}

The single crystal studied here was grown by radio-frequency skull
melting.\cite{butt95,rice93}  After orienting the crystal by X-ray Laue
diffraction, a  surface was cut perpendicular to the $c$ axis (parallel
to the NiO$_2$ planes)  and then polished.  Subsequently, the oxygen
concentration was selected by  annealing at 464$^\circ$C for 5 days in
flowing O$_2$ (1 bar), followed by a  quench to room temperature.  The
annealing conditions were chosen to give a  nominal $\delta$ of 0.133,
based on earlier work on the phase diagram.\cite{rice93}  A measurement
of the magnetization vs.\ temperature, with the  field applied parallel
to the planes, yielded the characteristic ferrimagnetic  response on
cooling, with an abrupt drop at the magnetic ordering temperature  of
110~K, observed in a previous study of the $\delta=\frac2{15}$
phase.\cite{tran97b}  

\begin{figure}[t]
\centerline{\includegraphics[width=3.2in]{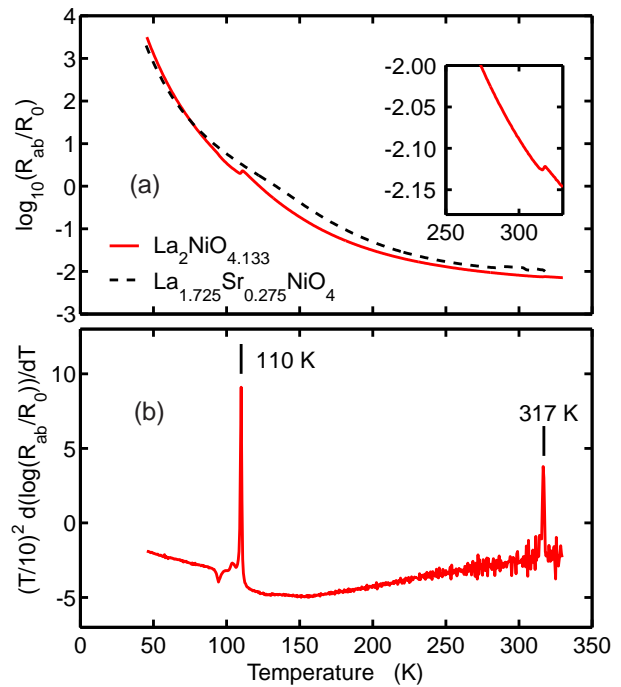}}
\medskip
\caption{(a)  Solid line: Logarithm of in-plane resistance divided by
$R_0 = 200~\Omega$ vs.\ temperature for La$_2$NiO$_{4.133}$ sample (inset
shows high-temperature range on an expanded scale); Dashed line:
Logarithm of in-plane resistivity (in units of $\Omega$~cm) for
La$_{1.725}$Sr$_{0.275}$NiO$_4$, from Ref.~\protect\onlinecite{lee02}. 
(b) Derivative of $\log(R_{ab}/R_0)$ with respect to temperature, weighted
by $(T/10)^2$ for the oxygen-doped sample.} \label{fg:rho} 
\end{figure}

The in-plane resistance, measured by the four-probe method on a small
slice of crystal, is plotted in Fig.~\ref{fg:rho}.   The effective
cross-sectional area, required for the conversion to resistivity,  could
not be determined because of the presence of micro-cracks, observed by 
optical microscopy, that result from quenching.  In an attempt to
compensate for this problem, we have plotted in the same figure the
in-plane resistivity of La$_{1.725}$Sr$_{0.275}$NiO$_4$, a sample with a
similar nominal hole concentration, normalized to $\rho_0=1\ \Omega$ cm. 
The normalization factor for the oxygen-doped crystal, $R_0 = 200\
\Omega$, has been chosen so that the two curves nearly coincide.

We expect the resistivities for the two samples, $\delta=\frac2{15}$ and
$x=0.275$, should be very similar since the hole density for each is
$n_h=x+2\delta\approx0.27$, and there is clearly little difference
between the temperature dependences.  We will use this similarity to
estimate the resistivity of the oxygen-doped sample:
$\rho_{ab} =\rho_0(R_{ab}/R_0)$, with $\rho_0 = 1\ \Omega$~cm.  This
estimated resistivity will be used in Sec.~\ref{sec:results}.

\begin{figure}[t]
\centerline{\includegraphics[width=3.2in]{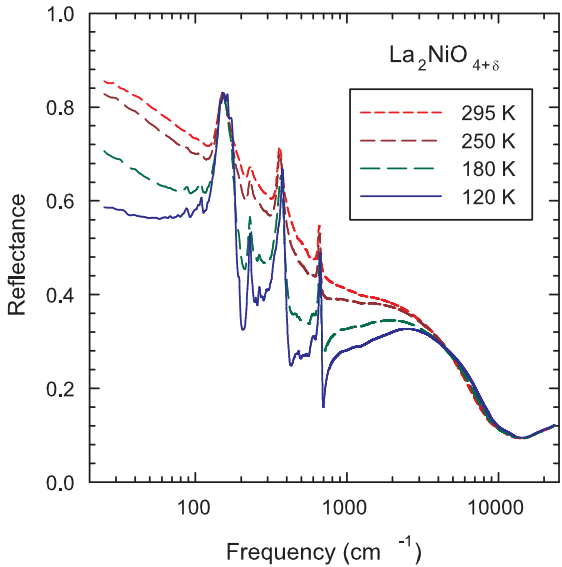}}
\medskip
\caption{Frequency dependence of infrared reflectance measured with
polarization parallel to the NiO$_2$ planes in La$_2$NiO$_{4.133}$.}
\label{fg:refl} 
\end{figure}

\begin{figure}[t]
\centerline{\includegraphics[width=3.2in]{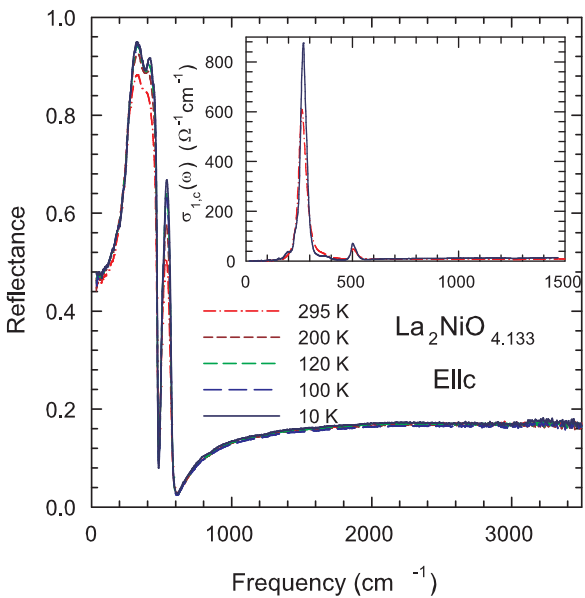}}
\medskip
\caption{Frequency dependence of infrared reflectance measured with
polarization perpendicular to the NiO$_2$ planes in La$_2$NiO$_{4.133}$.}
\label{fg:reflc} 
\end{figure}

For the IR measurements, the crystal was mounted in a cryostat on an 
optically-black cone. The temperature dependence of the reflectance was 
measured at a near-normal angle of incidence from $\approx 30$ to over 
24,000~cm$^{-1}$ on a Bruker IFS 66v/S using an {\it in situ} overcoating 
technique, which has previously been described in detail
elsewhere.\cite{home93}  Examples of reflectance measurements with the
light polarization perpendicular to and parallel to the $c$-axis are
shown in Figs.~\ref{fg:refl} and \ref{fg:reflc}, respectively.  For the
in-plane polarization, little temperature dependence is observed above
$\sim4000$~cm$^{-1}$, and hence the reflectance was assumed to be 
temperature independent for $\omega > 8000$~cm$^{-1}$. The optical
conductivity has been determined from a  Kramers-Kronig analysis of the
reflectance, for which extrapolations are  necessary for
$\omega\rightarrow 0, \infty$.  For the in-plane polarization, a metallic
low-frequency extrapolation, $R \propto 1-\omega^{1/2}$, was used for
$T\gtrsim 180$~K; at  lower temperatures the reflectance was assumed to
continue smoothly to $\approx  0.55$ at zero frequency.  

\section{Nature of Ordering in La$_2$NiO$_{4.133}$}

The interstitial oxygen order and the spin and charge stripe order of
\lnod\ with $\delta=\frac2{15}$ have been characterized previously by
neutron diffraction.\cite{tran94a,tran95b,woch98} The first
studies\cite{tran94a,tran95b} were done on a crystal with a net oxygen
excess of $\delta=0.125$, where the interstitial order developed below
$\sim310$~K.  The magnetic order appears abruptly below a first-order
transition\cite{tran94a,woch98} at 110~K.  Measurements on a crystal with
$\delta=0.133$ showed that the charge order survives well above 110~K, but
decays very slowly.\cite{woch98}  The intensity of a charge-order peak was
followed up to 220~K and was found to decay as $\exp(-T/T_0)$, similar to
a Debye-Waller factor.  

A piece of the present sample was studied by X-ray
diffraction.\cite{woch01} This time it was possible to follow the
gradual decay of charge order scattering up to $\approx300$~K before it
disappeared into the background; the exponential decrease of intensity
with temperature makes it difficult to evaluate where the charge-stripe
order truly goes to zero.  The interstitial order was found to disappear
at 317~K.

The interstitial order lowers the symmetry of the lattice, and provides
a modulated potential that can pin charge stripes.\cite{woch98}  The
absence of a clear disordering transition for the charge stripes is
consistent with order induced by a symmetry-breaking field, as occurs
in the case of a ferromagnet warmed in an applied magnetic field to
temperatures above its Curie point.  From such an analogy, it appears
possible that a finite degree of charge-stripe order appears as soon as
the interstitials order.  

To pursue this idea further, we have taken a close look at the in-plane
resistance.  There are two small jumps in the resistance, one at the
magnetic ordering temperature of 110~K and the second at 317~K where the
interstitials disorder.  (Note that the resistance {\em increases} on
warming through each of these transitions.)  To emphasize that these are
the only sharply defined transitions, we have plotted the derivative of
$\log(R_{ab}/R_0)$ with respect to temperature in Fig.~\ref{fg:rho}(b),
with the derivative multiplied by $T^2$ to make the higher temperature
feature clear.  The absence of a third transition is consistent with
simultaneous disordering of charge stripes and interstitials.

In \lsno, the highest charge ordering temperature, $T_{\rm co}$, is 240~K
at $x=\frac13$.  At a Sr concentration of $x=0.275$, closer to the hole
concentration of our sample, $T_{\rm co}\approx190$~K.  Thus, the likely
appearance of charge order at 317~K in the oxygen-doped crystal is quite
unusual.  Pinning of stripes by the interstitial potential suggests that
the stripes would exist dynamically in the absence of the potential. 
Direct evidence of dynamic stripes in a \lsno\ sample was recently
obtained in a neutron scattering study.\cite{lee02}

\begin{figure}[t]
\centerline{\includegraphics[width=3.2in]{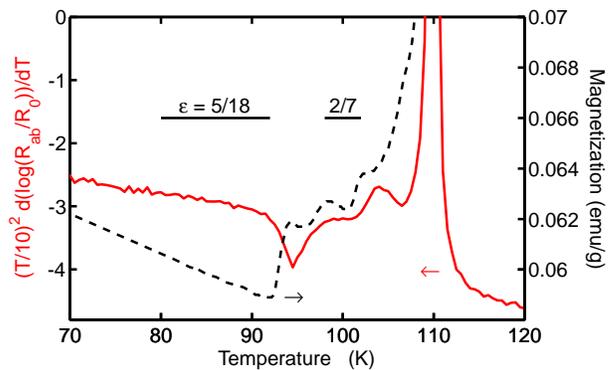}}
\medskip
\caption{Solid line: expanded view of $(T/10)^2 d(\log(R_{ab}/R_0))/dT$
from Fig.~\protect\ref{fg:rho}(b) in the region below the magnetic
ordering transition at 110~K.  Dashed line: magnetization measured on
another piece of the same sample with an applied field of 1~T.  Horizontal
bars indicate temperature ranges of lock-in plateaus, labelled with the
magnetic incommensurability, measured on a similar sample by neutron
diffraction.\protect\cite{woch98}}
\label{fg:drdt} 
\end{figure}

The attentive reader may have noticed the structure that appears in the
derivative of the resistivity below the 110~K spike in
Fig.~\ref{fg:rho}(b).  We have expanded this region in
Fig.~\ref{fg:drdt}, where we have also plotted the magnetization.  The
steps in the magnetization, which have previously been shown to
correspond to lock-in plateaus of the temperature-dependent inverse
stripe spacing,\cite{tran97b} also appear to correlate with changes in
the resistivity.  The resistivity is clearly sensitive to
variations in the stripe order.

\section{Expectations for Optical Conductivity due to Charge Stripes}
\label{sec:expect}

Before continuing on, let us discuss the possible interpretations of the
MIR  peak.  In \lsno\ with $x=0$, there is a charge-transfer-excitation
gap\cite{ido91,pell96} of 4~eV.  Doping introduces states within that
gap.  At room  temperature, the dopant-induced conductivity peak grows
with the Sr concentration but does not shift.  Integrating the
conductivity from 0 to 2~eV  yields an effective carrier concentration
equal to the Sr concentration  (assuming carriers have the free electron
mass).\cite{ido91}  Attempts to fit  the MIR peak with the conductivity
expected from a small polaron generally give  a poor overall fit to the
energy and temperature dependence.\cite{bi93,cran93,kats96}
Alternatively, Katsufuji {\it et al.}\cite{kats96} extracted a
temperature dependent gap from an analysis of the leading edge of the
MIR peak, and compared it with the temperature dependence of a CDW gap. 
Of  course, the MIR peak is still present when the putative gap goes to
zero, and the system never becomes a truly metallic conductor, so it is
difficult to see how a conventional CDW picture could be applied.

\begin{figure}[t]
\centerline{\includegraphics[width=3.4in]{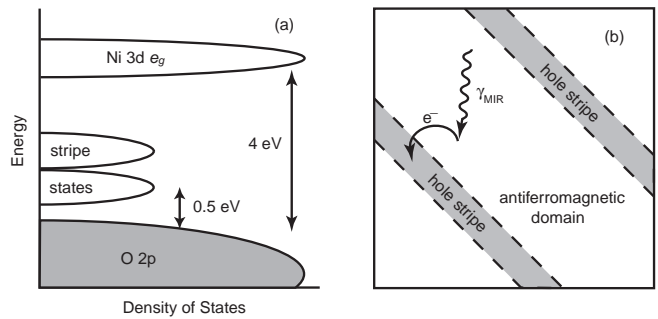}}
\caption{(a) Schematic diagram of the electronic density of states.  (b)
Cartoon illustrating the proposed absorption mechanism associated with
the MIR peak, in which an electron from an antiferromagnetic domain is
excited into an empty state on a hole stripe.}
\label{fg:dos}
\end{figure}

Any self-consistent interpretation of the MIR peak must take into account
the  fact that the doped holes tend to order as stripes.  Furthermore,
the stripe  solid melts into a stripe liquid above the charge-ordering
temperature,\cite{lee02} and so should still have observable effects on
the charge  excitations at higher temperatures.  Given the topological
nature of the observed charge stripes, one expects the electronic states
associated with them  to occur within the charge transfer
gap,\cite{zaan89,poil89,schu90,zaan96b,tche00,gran02,lore02} similar to
the soliton states observed in doped polyacetylene.\cite{heeg88} 
Consistent with this concept, it has been observed experimentally that
the chemical potential is pinned within the Mott-Hubbard gap for a
substantial part of the doping range over which stripe order is
observed.\cite{sata00} In the case of nickelates,  the mid-gap stripe
states should be empty \cite{zaan94}; Fig.~\ref{fg:dos}(a) shows a
schematic density of states.   In models appropriate to cuprates, the 
stripe states are quarter-filled with electrons.\cite{gran02}  Numerical 
calculations of optical conductivity for cuprate stripe models (without
electron-phonon coupling) yield both a Drude component and one or more MIR
peaks.\cite{verg91,salk96,mach99,shib01,mora02,lore02b}  (A related
result has been obtained by Caprara {\it et al.}\cite{capr02} using a
model of charge collective modes close to a charge-ordering instability.) 
We would attribute the Drude peak to excitations within the
partially-filled mid-gap states. In nickelates at higher temperatures it
may be possible to thermally excite carriers into the mid-gap states,
resulting in a relatively weak Drude peak.

%
%

But what about electron-phonon coupling?  Static atomic displacements are 
present in the charge-ordered state,\cite{tran95b} and these presumably
occur to lower the energy.  An estimate of the energy associated with
the lattice displacements is provided by an LDA$+U$ calculation by
Anisimov {\it et al.}\cite{anis92} for a nickelate with a hole
concentration of 0.25.  Assuming a  breathing-mode pattern of in-plane
oxygens about the 1/4 of the Ni sites with an extra hole, with a
displacement equal to 4\%\ of the bond length, the energy  gain is
210~meV.  When an electron is excited from the valence band in an
antiferromagnetic domain into a mid-gap stripe state [see
Fig.~\ref{fg:dos}(b)], the lattice does not have time to relax.  As a
result, the required excitation energy must be increased beyond the
minimum electronic energy by twice the lattice-relaxation energy: once in
removing the electron from the antiferromagnetic region and again in
adding the electron to a charge stripe.  This increase in the excitation
energy is known as the Franck-Condon effect.  We would expect this
``polaronic'' energy to  appear as a shift of the excitations in the
ordered state relative to that in the stripe-liquid state. 

If the theoretical estimate of a 4\%\ bond-length relaxation were
correct, then we might expect to see a temperature-dependent shift of
0.42~eV.  However, the bond-length modulation determined by a fit to
superlattice intensities\cite{tran95b} is approximately 2\%.  Since the
relaxation energy is proportional to the square of the displacement, a
better estimate of the shift is 0.1~eV.  As we will see below, the latter
value is roughly consistent with experiment.

\section{Optical Conductivity Results and Analysis}
\label{sec:results}

\begin{figure}[t]
\centerline{\includegraphics[width=3.2in]{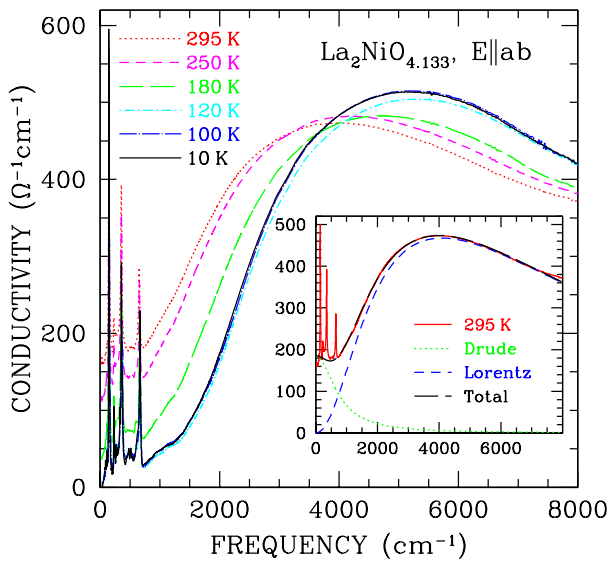}}
\medskip
\caption{The real part of the optical conductivity of La$_2$NiO$_{4.133}$
for  light polarized in the {\it ab} planes at a variety of temperatures.
Inset: Fit  to the conductivity at 295~K, and the decomposition of the
electronic  contribution into a Drude and MIR component.} \label{fg:1} 
\end{figure}

Figure~\ref{fg:1} shows the real part of the in-plane optical
conductivity. Beyond the sharp phonon peaks that appear below
700~cm$^{-1}$, one can see that the electronic conductivity is dominated
by the MIR peak at $\sim 5000\mbox{\rm\  cm}^{-1}\approx 0.6$~eV.  The
peak shifts to higher energy on cooling from 295~K to 120~K, with no
significant change below the latter temperature.  The  results are
qualitatively consistent with previous
results,\cite{kats96,pash00,poir02} although the size of the shift is
closer to that  found in \lsno\ with $x=0.225$ (Ref.~\onlinecite{pash00})
than that with $x=0.33$ (Ref.~\onlinecite{kats96}).

%
%

To quantitatively analyze the data, we have fit the conductivity
using a  Drude-Lorentz model for the dielectric function:
\begin{equation}
   \tilde\epsilon(\omega)=\epsilon_\infty-
   {{\omega_{p,D}^2} \over {\omega(\omega+i\Gamma_D)} }+
   \sum_j {{\omega_{p,j}^2} \over {\omega_j^2-\omega^2-i\gamma_j\omega}},
   \label{lorentz}
\end{equation}
where $\omega_{p,D}$ and $\Gamma_D$ are the classical plasma frequency
and  damping of the Drude (zero-frequency) component, and $\omega_j$,
$\gamma_j$ and $\omega_{p,j}$ are the frequency, width and effective
plasma frequency of the $j^{\rm th}$ vibration or electronic component;
$\epsilon_\infty$ is the core  contribution to the dielectric function.   
(A similar decomposition was used by Bi and Eklund at room
temperature.\cite{bi93}) The real part of the conductivity is  given by 
$\sigma_1=\omega{\rm Im}\,\tilde\epsilon/(4\pi)$.  The conductivity 
amplitude of the $j^{\rm th}$ contribution is 
$\sigma_1(\omega_j)=\omega_{p,j}^2/(4\pi\gamma_j)$.  Four strong phonon
modes were included in the model.

An example of fitted Drude and MIR components is presented in the inset of
Fig.~\ref{fg:1} for $T=295$~K; here $\omega_{p,D}\approx2800$~cm$^{-1}$
and $\Gamma_D\approx800$~cm$^{-1}$.  We find that the electronic
conductivity is well described by these two contributions at all
temperatures.  Now, the MIR peak should involve a convolution of
densities of states, and there is no particular reason to expect it to be
fitted well by a single Lorentz oscillator.  The fact that a single
oscillator provides a good fit is a convenient coincidence that
simplifies the following analysis.  We interpret the width as a measure 
of the convolved densities of occupied and unoccupied states expected for
an interband transition.

\begin{figure}[t]
\centerline{\includegraphics[width=3.0in]{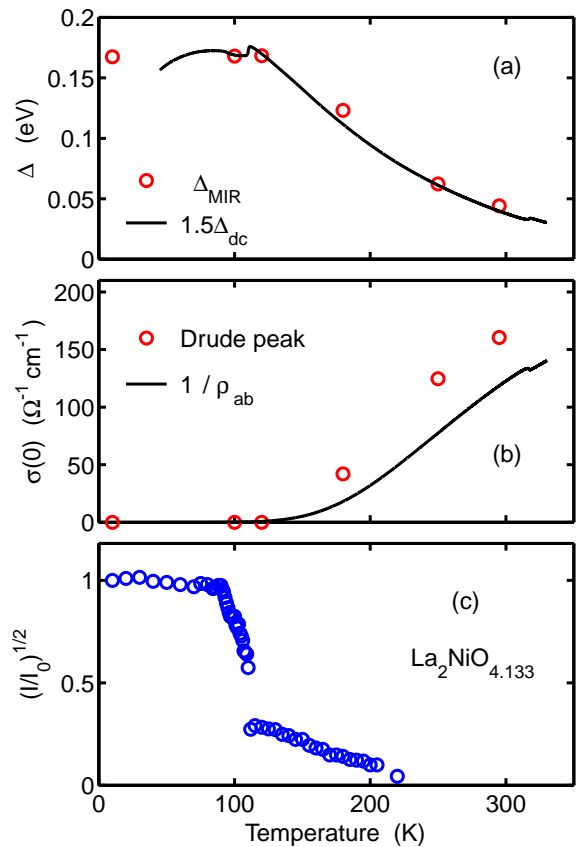}}
\caption{Temperature dependence of various parameters. (a) Energy shift
of the  MIR peak (circles), compared with the temperature dependence of
$\Delta_{dc}$  (Eq.~3) scaled to match the optical data. (b) 
Conductivity amplitude of Drude  component (circles), compared with the
inverse of the dc resistivity.  (c) Charge order parameter (square-root of
superlattice intensity) from neutron diffraction.\protect\cite{woch98}}
\label{fg:2} 
\end{figure}

The temperature dependence of the MIR peak energy is plotted in
Fig.~\ref{fg:2}(a) in the form $\Delta_{\rm MIR} = \hbar\omega_{\rm
MIR}(T)-\hbar\omega_0$, with $\hbar\omega_0 = 0.5$~eV.  The size of the 
MIR peak shift, 0.1--0.15~eV, is quite comparable to our estimate of the
Franck-Condon shift, 0.1~eV, based on the lattice relaxation energy
calculated by Anisimov {\it et al.}\cite{anis92}  One might expect such a
shift to create an optical gap $\Delta_{\rm dc}$, with 
$\Delta_{\rm dc}\sim\Delta_{\rm MIR}$.  Thermal
excitations of charge carriers into the mid-gap states would be limited
by this minimum gap; excitations at the MIR peak frequency are not
relevant for thermal processes.  If a gap of order
$\Delta_{\rm MIR}$ is present at low temperature in $\sigma(\omega)$, it
is masked by the phonon conductivity; nevertheless, the Drude conductivity
and the dc resistivity should be sensitive to such a gap.

The Drude conductivity is shown in Fig.~\ref{fg:2}(b).  On warming, it
rises from  zero as the MIR peak shifts to lower energy.  The solid line
in Fig.~\ref{fg:2}(b) represents the inverse of
$\rho_{ab}$.  The temperature dependence of the dc conductivity (inverse
resistivity) is consistent with the fitted Drude component. Assuming that
the Drude conductivity arises from thermal excitation of carriers across
$\Delta_{\rm dc}$, we expect the dc resistivity to behave roughly as 
\begin{equation}
   \rho_{ab} \approx \rho_0 e^{\Delta_{\rm dc}/2k_BT},
\end{equation}
so that
\begin{equation}
   \Delta_{\rm dc} = 2k_BT\ln(\rho_{ab}/\rho_0).
\end{equation}
The solid line in Fig.~\ref{fg:2}(a) shows the temperature dependence of
$\Delta_{\rm dc}$ obtained from $\rho_{ab}/\rho_0$, where we have set
$\rho_0=5$~m$\Omega$~cm, a value somewhat lower than the minimum observed
resistivity (see Fig.~\ref{fg:rho}). When scaled by a factor of 1.5,
$\Delta_{\rm dc}(T)$ agrees well with $\Delta_{\rm MIR}(T)$.  

%
%

By choosing a plausible $\rho_0$, we have extracted from the resistivity a
gap $\Delta_{\rm dc}(T)$ that is comparable to $\Delta_{\rm MIR}(T)$. 
This is consistent with our argument that the shift 
$\Delta_{\rm MIR}(T)$ results from the Franck-Condon effect.  But how does
the temperature dependence of the energy shift compare with the actual
lattice displacements?

The intensity of a charge-order superlattice peak measured by neutron
diffraction is proportional to the square of the lattice displacement. 
In Fig.~\ref{fg:2}(c) we have plotted the square-root of such an
intensity measurement.\cite{woch98}  (The choice of taking the
square-root was made to enhance the signal at higher temperatures.)  The
gradual decrease in the shift energy above the 110-K transition is
qualitatively consistent with the decay in magnitude of the lattice
displacements.  The contrasting behavior observed by Katsufuji {\it et
al.}\cite{kats96} in La$_{1.67}$Sr$_{0.33}$NiO$_4$, where the
energy-shift was found to exhibit a BCS-like temperature dependence,
corresponds with the critical behavior observed in the superlattice
intensities by  diffraction studies on that compound.\cite{lee97,du00} 

The feature in Fig.~\ref{fg:2} that, on first glance, seems inconsistent
is the jump in superlattice intensity at the 110-K spin ordering
transition---it is certainly not reflected in the optical energy shift. 
The discrepancy becomes less significant when one realizes that local
lattice displacements are not the only factor involved in stabilizing the
charge stripes.  At the first-order 110-K transition, there is a jump in
the ordering wave vector at the  transition; the charge stripes are
commensurate with the interstitial lattice above the transition, but
become effectively incommensurate below.\cite{woch98}  In the
commensurate phase, the stripes are stabilized by alignment with the
Coulomb potential of the interstitials, in addition to the local lattice
relaxation.  In the incommensurate phase at lower temperatures, the
Coulomb stabilization is lost, but the incommensurate phase is favored by
the reduction in energy from the ordering of the magnetic moments.  In
terms of screening the charge stripes, the loss of the interstitial
Coulomb energy is balanced by the increase in lattice displacements.
Thus, it seems reasonable that the energy shift of the MIR peak is
insensitive to the transition at 110~K. 

The magnitude of the MIR peak shift we observe is comparable to that
found in \lsno\ with $x=0.225$ (Ref.~\onlinecite{pash00}) and $0.5$
(Ref.~\onlinecite{jung01}), where the stripe order is
incommensurate.\cite{tran96a,kaji02}  In contrast, the shift is larger
(0.26 eV) for the $x=\frac13$ composition, where the order is
commensurate.\cite{kats96}  In the commensurate case, it seems likely
that the lattice displacements and the Coulomb energy work together,
resulting in an enhanced Franck-Condon energy.

Finally, we note that the electronic dynamical conductivity along the $c$
axis (see inset of Fig.~\ref{fg:reflc}) remains negligible at all
temperatures.  This result provides credibility to our assumption that
the electronic excitations probed with $\mathbf{\varepsilon}\perp{\bf c}$
are confined within the two-dimensional NiO$_2$ planes.  The insulating
behavior along the $c$-axis together with finite conductivity within the
planes is reminiscent of cuprate superconductors in the pseudogap
regime.\cite{home93b}

\section{Summary and Discussion}

We have presented a study of the optical conductivity in
La$_2$NiO$_{4.133}$.  While the data are qualitatively similar to those
reported for \lsno, we have introduced a new interpretation of the
electronic conductivity.  In particular, we have shown that
$\sigma_{ab}(\omega)$ can be decomposed into the MIR peak, a
contribution that is always gapped, and a Drude component that appears at
higher temperatures.  The two components are consistent with theoretical
expectations for the electronic structures associated wiht a
stripe-ordered system.  Associated with the charge stripes are mid-gap
states that, in the case of nickelates, are empty at low temperature. 
Transitions from the filled valence band to the empty mid-gap states
yield the MIR peak.  Thermal excitation of carriers into the mid-gap
states at higher temperature results in the Drude peak.  The
temperature-dependent frequency shift of the MIR peak is consistent with
a Franck-Condon effect that is dependent on the degree of stripe order.

The electronic structure associated with stripes is inhomogeneous in
real space. Thus, the MIR excitation is likely to involve moving an
electron from an antiferromagnetic region into a hole stripe, as
indicated in Fig.~\ref{fg:dos}(b). Thinking in terms of holes, the
transition involves a hole hopping off of a stripe, forming a local
fluctuation of that stripe.  The connection between stripe fluctuations
and conductivity (or hole motion) has been explored in a number of
theoretical studies of stripe models relevant to the
cuprates.\cite{tche00,zaan01,cher00,lore02b}  For the insulating stripes
in the nickelates, finite conductivity correlates with disordering of the
stripes.

The charge stripes in cuprates are partially filled with electrons, so
that Drude-like conduction should be possible even when the stripes are
ordered.  This appears to be the case in La$_{2-x-y}$Nd$_y$Sr$_x$CuO$_4$
based on transport\cite{noda99} and IR\cite{taji99,dumm02} studies.
Comparisons of the MIR conductivity in the nickelates with that
in cuprates have been made previously.\cite{bi90,ido91,bi93}  A
two-component  analysis (Drude$+$MIR), though controversial, has been
applied to measurements on cuprates where stripes have not been directly
detected, such as Bi$_2$Sr$_2$CaCu$_2$O$_8$ and Bi$_2$Sr$_2$CuO$_6$
(Ref.~\onlinecite{rome92,lupi00}).  We note that the values of $\Gamma_D$
at room temperature for those samples, 480~cm$^{-1}$ and 690~cm$^{-1}$
respectively, are not greatly different from the value of 800~cm$^{-1}$
for our nickelate sample.  On a related point, Ando {\it et
al.}\cite{ando01} have shown that in \lsco, where stripes are believed
to be relevant, the charge mobility at a given temperature changes by only
a very modest amount from the underdoped to optimally doped regimes, and
that it is quite similar in YBa$_2$Cu$_3$O$_{6+x}$.  The stripe-based
interpretation is certainly not the only one that one can imagine
applying to the cuprates, but the similarities between measurements on
cuprates and on the stripe-ordered nickelates lend considerable support
to this interpretation.

%
%

\section*{Acknowledgements}

We would like to thank S. A.~Kivelson, V. Perebeinos, D. N.~Basov,
M.~Strongin, S.~Brazovskii, A. L.~Chernyshev, R.~Werner, T. M. Rice, S.
Uchida, and J. Zaanen for helpful discussions.   Research at Brookhaven
is  supported by the Department of Energy's (DOE) Office  of Science
under Contract No.\ DE-AC02-98CH10886.  DJB acknowledges support  from
DOE under Contract No.\ DE-FG02-00ER45800. 


\begin{thebibliography}{57}
\expandafter\ifx\csname natexlab\endcsname\relax\def\natexlab#1{#1}\fi
\expandafter\ifx\csname bibnamefont\endcsname\relax
  \def\bibnamefont#1{#1}\fi
\expandafter\ifx\csname bibfnamefont\endcsname\relax
  \def\bibfnamefont#1{#1}\fi
\expandafter\ifx\csname citenamefont\endcsname\relax
  \def\citenamefont#1{#1}\fi
\expandafter\ifx\csname url\endcsname\relax
  \def\url#1{\texttt{#1}}\fi
\expandafter\ifx\csname urlprefix\endcsname\relax\def\urlprefix{URL }\fi
\providecommand{\bibinfo}[2]{#2}
\providecommand{\eprint}[2][]{\url{#2}}

\bibitem[{\citenamefont{Bi et~al.}(1990)\citenamefont{Bi, Eklund, McRae, Zhang,
  Metcalf, Spalek, and Honig}}]{bi90}
\bibinfo{author}{\bibfnamefont{X.-X.} \bibnamefont{Bi}},
  \bibinfo{author}{\bibfnamefont{P.~C.} \bibnamefont{Eklund}},
  \bibinfo{author}{\bibfnamefont{E.}~\bibnamefont{McRae}},
  \bibinfo{author}{\bibfnamefont{J.-G.} \bibnamefont{Zhang}},
  \bibinfo{author}{\bibfnamefont{P.}~\bibnamefont{Metcalf}},
  \bibinfo{author}{\bibfnamefont{J.}~\bibnamefont{Spalek}}, \bibnamefont{and}
  \bibinfo{author}{\bibfnamefont{J.~M.} \bibnamefont{Honig}},
  \bibinfo{journal}{Phys. Rev. B} \textbf{\bibinfo{volume}{42}},
  \bibinfo{pages}{4756} (\bibinfo{year}{1990}).

\bibitem[{\citenamefont{Ido et~al.}(1991)\citenamefont{Ido, Magoshi, Eisaki,
  and Uchida}}]{ido91}
\bibinfo{author}{\bibfnamefont{T.}~\bibnamefont{Ido}},
  \bibinfo{author}{\bibfnamefont{K.}~\bibnamefont{Magoshi}},
  \bibinfo{author}{\bibfnamefont{H.}~\bibnamefont{Eisaki}}, \bibnamefont{and}
  \bibinfo{author}{\bibfnamefont{S.}~\bibnamefont{Uchida}},
  \bibinfo{journal}{Phys. Rev. B} \textbf{\bibinfo{volume}{44}},
  \bibinfo{pages}{12094} (\bibinfo{year}{1991}).

\bibitem[{\citenamefont{Katsufuji et~al.}(1996)\citenamefont{Katsufuji, Tanabe,
  Ishikawa, Fukuda, Arima, and Tokura}}]{kats96}
\bibinfo{author}{\bibfnamefont{T.}~\bibnamefont{Katsufuji}},
  \bibinfo{author}{\bibfnamefont{T.}~\bibnamefont{Tanabe}},
  \bibinfo{author}{\bibfnamefont{T.}~\bibnamefont{Ishikawa}},
  \bibinfo{author}{\bibfnamefont{Y.}~\bibnamefont{Fukuda}},
  \bibinfo{author}{\bibfnamefont{T.}~\bibnamefont{Arima}}, \bibnamefont{and}
  \bibinfo{author}{\bibfnamefont{Y.}~\bibnamefont{Tokura}},
  \bibinfo{journal}{Phys. Rev. B} \textbf{\bibinfo{volume}{54}},
  \bibinfo{pages}{R14230} (\bibinfo{year}{1996}).

\bibitem[{\citenamefont{Bi and Eklund}(1993)}]{bi93}
\bibinfo{author}{\bibfnamefont{X.-X.} \bibnamefont{Bi}} \bibnamefont{and}
  \bibinfo{author}{\bibfnamefont{P.~C.} \bibnamefont{Eklund}},
  \bibinfo{journal}{Phys. Rev. Lett.} \textbf{\bibinfo{volume}{70}},
  \bibinfo{pages}{2625} (\bibinfo{year}{1993}).

\bibitem[{\citenamefont{Eagles et~al.}(1995)\citenamefont{Eagles, Lobo, and
  Gervais}}]{eagl95}
\bibinfo{author}{\bibfnamefont{D.~M.} \bibnamefont{Eagles}},
  \bibinfo{author}{\bibfnamefont{R.~P. S.~M.} \bibnamefont{Lobo}},
  \bibnamefont{and} \bibinfo{author}{\bibfnamefont{F.}~\bibnamefont{Gervais}},
  \bibinfo{journal}{Phys. Rev. B} \textbf{\bibinfo{volume}{52}},
  \bibinfo{pages}{6440} (\bibinfo{year}{1995}).

\bibitem[{\citenamefont{Calvani et~al.}(1996)\citenamefont{Calvani, Paolone,
  Dore, Lupi, Maselli, Medaglia, and Cheong}}]{calv96}
\bibinfo{author}{\bibfnamefont{P.}~\bibnamefont{Calvani}},
  \bibinfo{author}{\bibfnamefont{A.}~\bibnamefont{Paolone}},
  \bibinfo{author}{\bibfnamefont{P.}~\bibnamefont{Dore}},
  \bibinfo{author}{\bibfnamefont{S.}~\bibnamefont{Lupi}},
  \bibinfo{author}{\bibfnamefont{P.}~\bibnamefont{Maselli}},
  \bibinfo{author}{\bibfnamefont{P.~G.} \bibnamefont{Medaglia}},
  \bibnamefont{and} \bibinfo{author}{\bibfnamefont{S.-W.}
  \bibnamefont{Cheong}}, \bibinfo{journal}{Phys. Rev. B}
  \textbf{\bibinfo{volume}{54}}, \bibinfo{pages}{R9592} (\bibinfo{year}{1996}).

\bibitem[{\citenamefont{B\"auml et~al.}(1998)\citenamefont{B\"auml, Wellein,
  and Fehske}}]{baum98}
\bibinfo{author}{\bibfnamefont{B.}~\bibnamefont{B\"auml}},
  \bibinfo{author}{\bibfnamefont{G.}~\bibnamefont{Wellein}}, \bibnamefont{and}
  \bibinfo{author}{\bibfnamefont{H.}~\bibnamefont{Fehske}},
  \bibinfo{journal}{Phys. Rev. B} \textbf{\bibinfo{volume}{58}},
  \bibinfo{pages}{3663} (\bibinfo{year}{1998}).

\bibitem[{\citenamefont{Jung et~al.}(2001)\citenamefont{Jung, Kim, Noh, Kim,
  Ri, Levett, Lees, Paul, and Balakrishnan}}]{jung01}
\bibinfo{author}{\bibfnamefont{J.~H.} \bibnamefont{Jung}},
  \bibinfo{author}{\bibfnamefont{D.-W.} \bibnamefont{Kim}},
  \bibinfo{author}{\bibfnamefont{T.~W.} \bibnamefont{Noh}},
  \bibinfo{author}{\bibfnamefont{H.~C.} \bibnamefont{Kim}},
  \bibinfo{author}{\bibfnamefont{H.-C.} \bibnamefont{Ri}},
  \bibinfo{author}{\bibfnamefont{S.~J.} \bibnamefont{Levett}},
  \bibinfo{author}{\bibfnamefont{M.~R.} \bibnamefont{Lees}},
  \bibinfo{author}{\bibfnamefont{D.~M.} \bibnamefont{Paul}}, \bibnamefont{and}
  \bibinfo{author}{\bibfnamefont{G.}~\bibnamefont{Balakrishnan}},
  \bibinfo{journal}{Phys. Rev. B} \textbf{\bibinfo{volume}{64}},
  \bibinfo{pages}{165106} (\bibinfo{year}{2001}).

\bibitem[{\citenamefont{Tsutsui et~al.}(1999)\citenamefont{Tsutsui, Koshibae,
  and Maekawa}}]{tsut99}
\bibinfo{author}{\bibfnamefont{K.}~\bibnamefont{Tsutsui}},
  \bibinfo{author}{\bibfnamefont{W.}~\bibnamefont{Koshibae}}, \bibnamefont{and}
  \bibinfo{author}{\bibfnamefont{S.}~\bibnamefont{Maekawa}},
  \bibinfo{journal}{Phys. Rev. B} \textbf{\bibinfo{volume}{59}},
  \bibinfo{pages}{9729} (\bibinfo{year}{1999}).

\bibitem[{\citenamefont{Crandles et~al.}(1993)\citenamefont{Crandles, Timusk,
  Garret, and Greedan}}]{cran93}
\bibinfo{author}{\bibfnamefont{D.~A.} \bibnamefont{Crandles}},
  \bibinfo{author}{\bibfnamefont{T.}~\bibnamefont{Timusk}},
  \bibinfo{author}{\bibfnamefont{J.~D.} \bibnamefont{Garret}},
  \bibnamefont{and} \bibinfo{author}{\bibfnamefont{J.~E.}
  \bibnamefont{Greedan}}, \bibinfo{journal}{Physica C}
  \textbf{\bibinfo{volume}{216}}, \bibinfo{pages}{94} (\bibinfo{year}{1993}).

\bibitem[{\citenamefont{Pashkevich et~al.}(2000)\citenamefont{Pashkevich,
  Blinkin, Gnezdilov, Tsapenko, Eremenko, Lemmens, Fischer, Grove, G\"underodt,
  Degiorgi et~al.}}]{pash00}
\bibinfo{author}{\bibfnamefont{Y.~G.} \bibnamefont{Pashkevich}},
  \bibinfo{author}{\bibfnamefont{V.~A.} \bibnamefont{Blinkin}},
  \bibinfo{author}{\bibfnamefont{V.~P.} \bibnamefont{Gnezdilov}},
  \bibinfo{author}{\bibfnamefont{V.~V.} \bibnamefont{Tsapenko}},
  \bibinfo{author}{\bibfnamefont{V.~V.} \bibnamefont{Eremenko}},
  \bibinfo{author}{\bibfnamefont{P.}~\bibnamefont{Lemmens}},
  \bibinfo{author}{\bibfnamefont{M.}~\bibnamefont{Fischer}},
  \bibinfo{author}{\bibfnamefont{M.}~\bibnamefont{Grove}},
  \bibinfo{author}{\bibfnamefont{G.}~\bibnamefont{G\"underodt}},
  \bibinfo{author}{\bibfnamefont{L.}~\bibnamefont{Degiorgi}},
  \bibnamefont{et~al.}, \bibinfo{journal}{Phys. Rev. Lett.}
  \textbf{\bibinfo{volume}{84}}, \bibinfo{pages}{3919} (\bibinfo{year}{2000}).

\bibitem[{\citenamefont{Katsufuji et~al.}(1999)\citenamefont{Katsufuji, Tanabe,
  Ishikawa, Yamanouchi, Tokura, Kakeshita, Kajimoto, and Yoshizawa}}]{kats99}
\bibinfo{author}{\bibfnamefont{T.}~\bibnamefont{Katsufuji}},
  \bibinfo{author}{\bibfnamefont{T.}~\bibnamefont{Tanabe}},
  \bibinfo{author}{\bibfnamefont{T.}~\bibnamefont{Ishikawa}},
  \bibinfo{author}{\bibfnamefont{S.}~\bibnamefont{Yamanouchi}},
  \bibinfo{author}{\bibfnamefont{Y.}~\bibnamefont{Tokura}},
  \bibinfo{author}{\bibfnamefont{T.}~\bibnamefont{Kakeshita}},
  \bibinfo{author}{\bibfnamefont{R.}~\bibnamefont{Kajimoto}}, \bibnamefont{and}
  \bibinfo{author}{\bibfnamefont{H.}~\bibnamefont{Yoshizawa}},
  \bibinfo{journal}{Phys. Rev. B} \textbf{\bibinfo{volume}{60}},
  \bibinfo{pages}{R5097} (\bibinfo{year}{1999}).

\bibitem[{\citenamefont{Lee and Cheong}(1997)}]{lee97}
\bibinfo{author}{\bibfnamefont{S.-H.} \bibnamefont{Lee}} \bibnamefont{and}
  \bibinfo{author}{\bibfnamefont{S.-W.} \bibnamefont{Cheong}},
  \bibinfo{journal}{Phys. Rev. Lett.} \textbf{\bibinfo{volume}{79}},
  \bibinfo{pages}{2514} (\bibinfo{year}{1997}).

\bibitem[{\citenamefont{Tranquada et~al.}(1994)\citenamefont{Tranquada,
  Buttrey, Sachan, and Lorenzo}}]{tran94a}
\bibinfo{author}{\bibfnamefont{J.~M.} \bibnamefont{Tranquada}},
  \bibinfo{author}{\bibfnamefont{D.~J.} \bibnamefont{Buttrey}},
  \bibinfo{author}{\bibfnamefont{V.}~\bibnamefont{Sachan}}, \bibnamefont{and}
  \bibinfo{author}{\bibfnamefont{J.~E.} \bibnamefont{Lorenzo}},
  \bibinfo{journal}{Phys. Rev. Lett.} \textbf{\bibinfo{volume}{73}},
  \bibinfo{pages}{1003} (\bibinfo{year}{1994}).

\bibitem[{\citenamefont{Tranquada}(1998)}]{tran98b}
\bibinfo{author}{\bibfnamefont{J.~M.} \bibnamefont{Tranquada}}, in
  \emph{\bibinfo{booktitle}{Neutron Scattering in Layered Copper-Oxide
  Superconductors}}, edited by
  \bibinfo{editor}{\bibfnamefont{A.}~\bibnamefont{Furrer}}
  (\bibinfo{publisher}{Kluwer}, \bibinfo{address}{Dordrecht, The Netherlands},
  \bibinfo{year}{1998}), p. \bibinfo{pages}{225}.

\bibitem[{\citenamefont{Yoshizawa et~al.}(2000)\citenamefont{Yoshizawa,
  Kakeshita, Kajimoto, Tanabe, Katsufuji, and Tokura}}]{yosh00}
\bibinfo{author}{\bibfnamefont{H.}~\bibnamefont{Yoshizawa}},
  \bibinfo{author}{\bibfnamefont{T.}~\bibnamefont{Kakeshita}},
  \bibinfo{author}{\bibfnamefont{R.}~\bibnamefont{Kajimoto}},
  \bibinfo{author}{\bibfnamefont{T.}~\bibnamefont{Tanabe}},
  \bibinfo{author}{\bibfnamefont{T.}~\bibnamefont{Katsufuji}},
  \bibnamefont{and} \bibinfo{author}{\bibfnamefont{Y.}~\bibnamefont{Tokura}},
  \bibinfo{journal}{Phys. Rev. B} \textbf{\bibinfo{volume}{61}},
  \bibinfo{pages}{R854} (\bibinfo{year}{2000}).

\bibitem[{\citenamefont{Lee et~al.}(2002)\citenamefont{Lee, Tranquada, Yamada,
  Buttrey, Li, and Cheong}}]{lee02}
\bibinfo{author}{\bibfnamefont{S.-H.} \bibnamefont{Lee}},
  \bibinfo{author}{\bibfnamefont{J.~M.} \bibnamefont{Tranquada}},
  \bibinfo{author}{\bibfnamefont{K.}~\bibnamefont{Yamada}},
  \bibinfo{author}{\bibfnamefont{D.~J.} \bibnamefont{Buttrey}},
  \bibinfo{author}{\bibfnamefont{Q.}~\bibnamefont{Li}}, \bibnamefont{and}
  \bibinfo{author}{\bibfnamefont{S.-W.} \bibnamefont{Cheong}},
  \bibinfo{journal}{Phys. Rev. Lett.} \textbf{\bibinfo{volume}{88}},
  \bibinfo{pages}{126401} (\bibinfo{year}{2002}).

\bibitem[{\citenamefont{Wochner et~al.}(1998)\citenamefont{Wochner, Tranquada,
  Buttrey, and Sachan}}]{woch98}
\bibinfo{author}{\bibfnamefont{P.}~\bibnamefont{Wochner}},
  \bibinfo{author}{\bibfnamefont{J.~M.} \bibnamefont{Tranquada}},
  \bibinfo{author}{\bibfnamefont{D.~J.} \bibnamefont{Buttrey}},
  \bibnamefont{and} \bibinfo{author}{\bibfnamefont{V.}~\bibnamefont{Sachan}},
  \bibinfo{journal}{Phys. Rev. B} \textbf{\bibinfo{volume}{57}},
  \bibinfo{pages}{1066} (\bibinfo{year}{1998}).

\bibitem[{\citenamefont{Tranquada et~al.}(1995)\citenamefont{Tranquada,
  Lorenzo, Buttrey, and Sachan}}]{tran95b}
\bibinfo{author}{\bibfnamefont{J.~M.} \bibnamefont{Tranquada}},
  \bibinfo{author}{\bibfnamefont{J.~E.} \bibnamefont{Lorenzo}},
  \bibinfo{author}{\bibfnamefont{D.~J.} \bibnamefont{Buttrey}},
  \bibnamefont{and} \bibinfo{author}{\bibfnamefont{V.}~\bibnamefont{Sachan}},
  \bibinfo{journal}{Phys. Rev. B} \textbf{\bibinfo{volume}{52}},
  \bibinfo{pages}{3581} (\bibinfo{year}{1995}).

\bibitem[{\citenamefont{Granath et~al.}(2002)\citenamefont{Granath, Oganesyan,
  Orgad, and Kivelson}}]{gran02}
\bibinfo{author}{\bibfnamefont{M.}~\bibnamefont{Granath}},
  \bibinfo{author}{\bibfnamefont{V.}~\bibnamefont{Oganesyan}},
  \bibinfo{author}{\bibfnamefont{D.}~\bibnamefont{Orgad}}, \bibnamefont{and}
  \bibinfo{author}{\bibfnamefont{S.~A.} \bibnamefont{Kivelson}},
  \bibinfo{journal}{Phys. Rev. B} \textbf{\bibinfo{volume}{65}},
  \bibinfo{pages}{184501} (\bibinfo{year}{2002}).

\bibitem[{\citenamefont{Heeger et~al.}(1988)\citenamefont{Heeger, Kivelson,
  Schrieffer, and Su}}]{heeg88}
\bibinfo{author}{\bibfnamefont{A.~J.} \bibnamefont{Heeger}},
  \bibinfo{author}{\bibfnamefont{S.}~\bibnamefont{Kivelson}},
  \bibinfo{author}{\bibfnamefont{J.~R.} \bibnamefont{Schrieffer}},
  \bibnamefont{and} \bibinfo{author}{\bibfnamefont{W.-P.} \bibnamefont{Su}},
  \bibinfo{journal}{Rev. Mod. Phys.} \textbf{\bibinfo{volume}{60}},
  \bibinfo{pages}{781} (\bibinfo{year}{1988}).

\bibitem[{\citenamefont{Tchernyshyov and Pryadko}(2000)}]{tche00}
\bibinfo{author}{\bibfnamefont{O.}~\bibnamefont{Tchernyshyov}}
  \bibnamefont{and} \bibinfo{author}{\bibfnamefont{L.~P.}
  \bibnamefont{Pryadko}}, \bibinfo{journal}{Phys. Rev. B}
  \textbf{\bibinfo{volume}{61}}, \bibinfo{pages}{12503} (\bibinfo{year}{2000}).

\bibitem[{\citenamefont{Zaanen and Gunnarsson}(1989)}]{zaan89}
\bibinfo{author}{\bibfnamefont{J.}~\bibnamefont{Zaanen}} \bibnamefont{and}
  \bibinfo{author}{\bibfnamefont{O.}~\bibnamefont{Gunnarsson}},
  \bibinfo{journal}{Phys. Rev. B} \textbf{\bibinfo{volume}{40}},
  \bibinfo{pages}{7391} (\bibinfo{year}{1989}).

\bibitem[{\citenamefont{Zaanen and Ole\'s}(1996)}]{zaan96b}
\bibinfo{author}{\bibfnamefont{J.}~\bibnamefont{Zaanen}} \bibnamefont{and}
  \bibinfo{author}{\bibfnamefont{A.~M.} \bibnamefont{Ole\'s}},
  \bibinfo{journal}{Ann. Phys. (Leipzig)} \textbf{\bibinfo{volume}{5}},
  \bibinfo{pages}{224} (\bibinfo{year}{1996}).

\bibitem[{\citenamefont{Poilblanc and Rice}(1989)}]{poil89}
\bibinfo{author}{\bibfnamefont{D.}~\bibnamefont{Poilblanc}} \bibnamefont{and}
  \bibinfo{author}{\bibfnamefont{T.~M.} \bibnamefont{Rice}},
  \bibinfo{journal}{Phys. Rev. B} \textbf{\bibinfo{volume}{39}},
  \bibinfo{pages}{9749} (\bibinfo{year}{1989}).

\bibitem[{\citenamefont{Schulz}(1990)}]{schu90}
\bibinfo{author}{\bibfnamefont{H.~J.} \bibnamefont{Schulz}},
  \bibinfo{journal}{Phys. Rev. Lett.} \textbf{\bibinfo{volume}{64}},
  \bibinfo{pages}{1445} (\bibinfo{year}{1990}).

\bibitem[{\citenamefont{Lorenzana and Seibold}(2002)}]{lore02}
\bibinfo{author}{\bibfnamefont{J.}~\bibnamefont{Lorenzana}} \bibnamefont{and}
  \bibinfo{author}{\bibfnamefont{G.}~\bibnamefont{Seibold}},
  \bibinfo{journal}{Phys. Rev. Lett.} \textbf{\bibinfo{volume}{89}},
  \bibinfo{pages}{136401} (\bibinfo{year}{2002}).

\bibitem[{\citenamefont{Ichikawa et~al.}(2000)\citenamefont{Ichikawa, Uchida,
  Tranquada, Niem\"oller, Gehring, Lee, and Schneider}}]{ichi00}
\bibinfo{author}{\bibfnamefont{N.}~\bibnamefont{Ichikawa}},
  \bibinfo{author}{\bibfnamefont{S.}~\bibnamefont{Uchida}},
  \bibinfo{author}{\bibfnamefont{J.~M.} \bibnamefont{Tranquada}},
  \bibinfo{author}{\bibfnamefont{T.}~\bibnamefont{Niem\"oller}},
  \bibinfo{author}{\bibfnamefont{P.~M.} \bibnamefont{Gehring}},
  \bibinfo{author}{\bibfnamefont{S.-H.} \bibnamefont{Lee}}, \bibnamefont{and}
  \bibinfo{author}{\bibfnamefont{J.~R.} \bibnamefont{Schneider}},
  \bibinfo{journal}{Phys. Rev. Lett.} \textbf{\bibinfo{volume}{85}},
  \bibinfo{pages}{1738} (\bibinfo{year}{2000}).

\bibitem[{\citenamefont{Tajima et~al.}(1999)\citenamefont{Tajima, Wang,
  Ichikawa, Eisaki, Uchida, Kitano, Hanaguri, and Maeda}}]{taji99}
\bibinfo{author}{\bibfnamefont{S.}~\bibnamefont{Tajima}},
  \bibinfo{author}{\bibfnamefont{N.~L.} \bibnamefont{Wang}},
  \bibinfo{author}{\bibfnamefont{N.}~\bibnamefont{Ichikawa}},
  \bibinfo{author}{\bibfnamefont{H.}~\bibnamefont{Eisaki}},
  \bibinfo{author}{\bibfnamefont{S.}~\bibnamefont{Uchida}},
  \bibinfo{author}{\bibfnamefont{H.}~\bibnamefont{Kitano}},
  \bibinfo{author}{\bibfnamefont{T.}~\bibnamefont{Hanaguri}}, \bibnamefont{and}
  \bibinfo{author}{\bibfnamefont{A.}~\bibnamefont{Maeda}},
  \bibinfo{journal}{Europhys. Lett.} \textbf{\bibinfo{volume}{47}},
  \bibinfo{pages}{715} (\bibinfo{year}{1999}).

\bibitem[{\citenamefont{Noda et~al.}(1999)\citenamefont{Noda, Eisaki, and
  Uchida}}]{noda99}
\bibinfo{author}{\bibfnamefont{T.}~\bibnamefont{Noda}},
  \bibinfo{author}{\bibfnamefont{H.}~\bibnamefont{Eisaki}}, \bibnamefont{and}
  \bibinfo{author}{\bibfnamefont{S.}~\bibnamefont{Uchida}},
  \bibinfo{journal}{Science} \textbf{\bibinfo{volume}{286}},
  \bibinfo{pages}{265} (\bibinfo{year}{1999}).

\bibitem[{\citenamefont{Dumm et~al.}(2002)\citenamefont{Dumm, Basov, Komiya,
  Abe, and Ando}}]{dumm02}
\bibinfo{author}{\bibfnamefont{M.}~\bibnamefont{Dumm}},
  \bibinfo{author}{\bibfnamefont{D.~N.} \bibnamefont{Basov}},
  \bibinfo{author}{\bibfnamefont{S.}~\bibnamefont{Komiya}},
  \bibinfo{author}{\bibfnamefont{Y.}~\bibnamefont{Abe}}, \bibnamefont{and}
  \bibinfo{author}{\bibfnamefont{Y.}~\bibnamefont{Ando}},
  \bibinfo{journal}{Phys. Rev. Lett.} \textbf{\bibinfo{volume}{88}},
  \bibinfo{pages}{147003} (\bibinfo{year}{2002}).

\bibitem[{\citenamefont{Rice and Buttrey}(1993)}]{rice93}
\bibinfo{author}{\bibfnamefont{D.~E.} \bibnamefont{Rice}} \bibnamefont{and}
  \bibinfo{author}{\bibfnamefont{D.~J.} \bibnamefont{Buttrey}},
  \bibinfo{journal}{J. Solid State Chem.} \textbf{\bibinfo{volume}{105}},
  \bibinfo{pages}{197} (\bibinfo{year}{1993}).

\bibitem[{\citenamefont{Buttrey et~al.}(1995)\citenamefont{Buttrey, Schartman,
  and Honig}}]{butt95}
\bibinfo{author}{\bibfnamefont{D.~J.} \bibnamefont{Buttrey}},
  \bibinfo{author}{\bibfnamefont{R.~R.} \bibnamefont{Schartman}},
  \bibnamefont{and} \bibinfo{author}{\bibfnamefont{J.~M.} \bibnamefont{Honig}},
  \bibinfo{journal}{Inorg. Synth.} \textbf{\bibinfo{volume}{30}},
  \bibinfo{pages}{130} (\bibinfo{year}{1995}).

\bibitem[{\citenamefont{Tranquada et~al.}(1997)\citenamefont{Tranquada,
  Wochner, Moodenbaugh, and Buttrey}}]{tran97b}
\bibinfo{author}{\bibfnamefont{J.~M.} \bibnamefont{Tranquada}},
  \bibinfo{author}{\bibfnamefont{P.}~\bibnamefont{Wochner}},
  \bibinfo{author}{\bibfnamefont{A.~R.} \bibnamefont{Moodenbaugh}},
  \bibnamefont{and} \bibinfo{author}{\bibfnamefont{D.~J.}
  \bibnamefont{Buttrey}}, \bibinfo{journal}{Phys. Rev. B}
  \textbf{\bibinfo{volume}{55}}, \bibinfo{pages}{R6113} (\bibinfo{year}{1997}).

\bibitem[{\citenamefont{Homes et~al.}(1993{\natexlab{a}})\citenamefont{Homes,
  Reedyk, Crandles, and Timusk}}]{home93}
\bibinfo{author}{\bibfnamefont{C.~C.} \bibnamefont{Homes}},
  \bibinfo{author}{\bibfnamefont{M.}~\bibnamefont{Reedyk}},
  \bibinfo{author}{\bibfnamefont{D.}~\bibnamefont{Crandles}}, \bibnamefont{and}
  \bibinfo{author}{\bibfnamefont{T.}~\bibnamefont{Timusk}},
  \bibinfo{journal}{Appl. Opt.} \textbf{\bibinfo{volume}{32}},
  \bibinfo{pages}{2972} (\bibinfo{year}{1993}{\natexlab{a}}).

\bibitem[{\citenamefont{Wochner et~al.}()\citenamefont{Wochner, Tranquada,
  Nelson, Hill, Gibbs, and Buttrey}}]{woch01}
\bibinfo{author}{\bibfnamefont{P.}~\bibnamefont{Wochner}},
  \bibinfo{author}{\bibfnamefont{J.~M.} \bibnamefont{Tranquada}},
  \bibinfo{author}{\bibfnamefont{C.~S.} \bibnamefont{Nelson}},
  \bibinfo{author}{\bibfnamefont{J.~P.} \bibnamefont{Hill}},
  \bibinfo{author}{\bibfnamefont{D.}~\bibnamefont{Gibbs}}, \bibnamefont{and}
  \bibinfo{author}{\bibfnamefont{D.~J.} \bibnamefont{Buttrey}},
  \bibinfo{note}{(unpublished)}.

\bibitem[{\citenamefont{Pellegrin et~al.}(1996)\citenamefont{Pellegrin, Zaanen,
  Lin, Meigs, Chen, Ho, Eisaki, and Uchida}}]{pell96}
\bibinfo{author}{\bibfnamefont{E.}~\bibnamefont{Pellegrin}},
  \bibinfo{author}{\bibfnamefont{J.}~\bibnamefont{Zaanen}},
  \bibinfo{author}{\bibfnamefont{H.-J.} \bibnamefont{Lin}},
  \bibinfo{author}{\bibfnamefont{G.}~\bibnamefont{Meigs}},
  \bibinfo{author}{\bibfnamefont{C.~T.} \bibnamefont{Chen}},
  \bibinfo{author}{\bibfnamefont{G.~H.} \bibnamefont{Ho}},
  \bibinfo{author}{\bibfnamefont{H.}~\bibnamefont{Eisaki}}, \bibnamefont{and}
  \bibinfo{author}{\bibfnamefont{S.}~\bibnamefont{Uchida}},
  \bibinfo{journal}{Phys. Rev. B} \textbf{\bibinfo{volume}{53}},
  \bibinfo{pages}{10667} (\bibinfo{year}{1996}).

\bibitem[{\citenamefont{Satake et~al.}(2000)\citenamefont{Satake, Kobayashi,
  Mizokawa, Fujimori, Tanabe, Katsufuji, and Tokura}}]{sata00}
\bibinfo{author}{\bibfnamefont{M.}~\bibnamefont{Satake}},
  \bibinfo{author}{\bibfnamefont{K.}~\bibnamefont{Kobayashi}},
  \bibinfo{author}{\bibfnamefont{T.}~\bibnamefont{Mizokawa}},
  \bibinfo{author}{\bibfnamefont{A.}~\bibnamefont{Fujimori}},
  \bibinfo{author}{\bibfnamefont{T.}~\bibnamefont{Tanabe}},
  \bibinfo{author}{\bibfnamefont{T.}~\bibnamefont{Katsufuji}},
  \bibnamefont{and} \bibinfo{author}{\bibfnamefont{Y.}~\bibnamefont{Tokura}},
  \bibinfo{journal}{Phys. Rev. B} \textbf{\bibinfo{volume}{61}},
  \bibinfo{pages}{15515} (\bibinfo{year}{2000}).

\bibitem[{\citenamefont{Zaanen and Littlewood}(1994)}]{zaan94}
\bibinfo{author}{\bibfnamefont{J.}~\bibnamefont{Zaanen}} \bibnamefont{and}
  \bibinfo{author}{\bibfnamefont{P.~B.} \bibnamefont{Littlewood}},
  \bibinfo{journal}{Phys. Rev. B} \textbf{\bibinfo{volume}{50}},
  \bibinfo{pages}{7222} (\bibinfo{year}{1994}).

\bibitem[{\citenamefont{Salkola et~al.}(1996)\citenamefont{Salkola, Emery, and
  Kivelson}}]{salk96}
\bibinfo{author}{\bibfnamefont{M.}~\bibnamefont{Salkola}},
  \bibinfo{author}{\bibfnamefont{V.~J.} \bibnamefont{Emery}}, \bibnamefont{and}
  \bibinfo{author}{\bibfnamefont{S.~A.} \bibnamefont{Kivelson}},
  \bibinfo{journal}{Phys. Rev. Lett.} \textbf{\bibinfo{volume}{77}},
  \bibinfo{pages}{155} (\bibinfo{year}{1996}).

\bibitem[{\citenamefont{Verg\'es et~al.}(1991)\citenamefont{Verg\'es, Louis,
  Lomdahl, Guinea, and Bishop}}]{verg91}
\bibinfo{author}{\bibfnamefont{J.~A.} \bibnamefont{Verg\'es}},
  \bibinfo{author}{\bibfnamefont{E.}~\bibnamefont{Louis}},
  \bibinfo{author}{\bibfnamefont{P.~S.} \bibnamefont{Lomdahl}},
  \bibinfo{author}{\bibfnamefont{F.}~\bibnamefont{Guinea}}, \bibnamefont{and}
  \bibinfo{author}{\bibfnamefont{A.~R.} \bibnamefont{Bishop}},
  \bibinfo{journal}{Phys. Rev. B} \textbf{\bibinfo{volume}{43}},
  \bibinfo{pages}{6099} (\bibinfo{year}{1991}).

\bibitem[{\citenamefont{Machida and Ichioka}(1999)}]{mach99}
\bibinfo{author}{\bibfnamefont{K.}~\bibnamefont{Machida}} \bibnamefont{and}
  \bibinfo{author}{\bibfnamefont{M.}~\bibnamefont{Ichioka}},
  \bibinfo{journal}{J. Phys. Soc. Japan} \textbf{\bibinfo{volume}{68}},
  \bibinfo{pages}{2168} (\bibinfo{year}{1999}).

\bibitem[{\citenamefont{Shibata et~al.}(2001)\citenamefont{Shibata, Tohyama,
  and Maekawa}}]{shib01}
\bibinfo{author}{\bibfnamefont{Y.}~\bibnamefont{Shibata}},
  \bibinfo{author}{\bibfnamefont{T.}~\bibnamefont{Tohyama}}, \bibnamefont{and}
  \bibinfo{author}{\bibfnamefont{S.}~\bibnamefont{Maekawa}},
  \bibinfo{journal}{Phys. Rev. B} \textbf{\bibinfo{volume}{64}},
  \bibinfo{pages}{054519} (\bibinfo{year}{2001}).

\bibitem[{\citenamefont{Moraghebi et~al.}(2002)\citenamefont{Moraghebi, Yunoki,
  and Moreo}}]{mora02}
\bibinfo{author}{\bibfnamefont{M.}~\bibnamefont{Moraghebi}},
  \bibinfo{author}{\bibfnamefont{S.}~\bibnamefont{Yunoki}}, \bibnamefont{and}
  \bibinfo{author}{\bibfnamefont{A.}~\bibnamefont{Moreo}},
  \bibinfo{journal}{Phys. Rev. B} \textbf{\bibinfo{volume}{66}},
  \bibinfo{pages}{214522} (\bibinfo{year}{2002}).

\bibitem[{\citenamefont{Lorenzana and Seibold}()}]{lore02b}
\bibinfo{author}{\bibfnamefont{J.}~\bibnamefont{Lorenzana}} \bibnamefont{and}
  \bibinfo{author}{\bibfnamefont{G.}~\bibnamefont{Seibold}},
  \bibinfo{journal}{Phys. Rev. Lett.} \textbf{\bibinfo{volume}{90}},
  \bibinfo{pages}{066404} (\bibinfo{year}{2003}).

\bibitem[{\citenamefont{Caprara et~al.}(2002)\citenamefont{Caprara, {Di
  Castro}, Fratini, and Grilli}}]{capr02}
\bibinfo{author}{\bibfnamefont{S.}~\bibnamefont{Caprara}},
  \bibinfo{author}{\bibfnamefont{C.}~\bibnamefont{{Di Castro}}},
  \bibinfo{author}{\bibfnamefont{S.}~\bibnamefont{Fratini}}, \bibnamefont{and}
  \bibinfo{author}{\bibfnamefont{M.}~\bibnamefont{Grilli}},
  \bibinfo{journal}{Phys. Rev. Lett.} \textbf{\bibinfo{volume}{88}},
  \bibinfo{pages}{147001} (\bibinfo{year}{2002}).

\bibitem[{\citenamefont{Anisimov et~al.}(1992)\citenamefont{Anisimov, Korotin,
  Zaanen, and Andersen}}]{anis92}
\bibinfo{author}{\bibfnamefont{V.~I.} \bibnamefont{Anisimov}},
  \bibinfo{author}{\bibfnamefont{M.~A.} \bibnamefont{Korotin}},
  \bibinfo{author}{\bibfnamefont{J.}~\bibnamefont{Zaanen}}, \bibnamefont{and}
  \bibinfo{author}{\bibfnamefont{O.~K.} \bibnamefont{Andersen}},
  \bibinfo{journal}{Phys. Rev. Lett.} \textbf{\bibinfo{volume}{68}},
  \bibinfo{pages}{345} (\bibinfo{year}{1992}).

\bibitem[{\citenamefont{Poirot-Reveau et~al.}(2002)\citenamefont{Poirot-Reveau,
  Odier, Simon, and Gervais}}]{poir02}
\bibinfo{author}{\bibfnamefont{N.}~\bibnamefont{Poirot-Reveau}},
  \bibinfo{author}{\bibfnamefont{P.}~\bibnamefont{Odier}},
  \bibinfo{author}{\bibfnamefont{P.}~\bibnamefont{Simon}}, \bibnamefont{and}
  \bibinfo{author}{\bibfnamefont{F.}~\bibnamefont{Gervais}},
  \bibinfo{journal}{Phys. Rev. B} \textbf{\bibinfo{volume}{65}},
  \bibinfo{pages}{094503} (\bibinfo{year}{2002}).

\bibitem[{\citenamefont{Du et~al.}(2000)\citenamefont{Du, Ghazi, Su, Pape,
  Hatton, Brown, Stirling, Cooper, and Cheong}}]{du00}
\bibinfo{author}{\bibfnamefont{C.-H.} \bibnamefont{Du}},
  \bibinfo{author}{\bibfnamefont{M.~E.} \bibnamefont{Ghazi}},
  \bibinfo{author}{\bibfnamefont{Y.}~\bibnamefont{Su}},
  \bibinfo{author}{\bibfnamefont{I.}~\bibnamefont{Pape}},
  \bibinfo{author}{\bibfnamefont{P.~D.} \bibnamefont{Hatton}},
  \bibinfo{author}{\bibfnamefont{S.~D.} \bibnamefont{Brown}},
  \bibinfo{author}{\bibfnamefont{W.~G.} \bibnamefont{Stirling}},
  \bibinfo{author}{\bibfnamefont{M.~J.} \bibnamefont{Cooper}},
  \bibnamefont{and} \bibinfo{author}{\bibfnamefont{S.-W.}
  \bibnamefont{Cheong}}, \bibinfo{journal}{Phys. Rev. Lett.}
  \textbf{\bibinfo{volume}{84}}, \bibinfo{pages}{3911} (\bibinfo{year}{2000}).

\bibitem[{\citenamefont{Tranquada et~al.}(1996)\citenamefont{Tranquada,
  Buttrey, and Sachan}}]{tran96a}
\bibinfo{author}{\bibfnamefont{J.~M.} \bibnamefont{Tranquada}},
  \bibinfo{author}{\bibfnamefont{D.~J.} \bibnamefont{Buttrey}},
  \bibnamefont{and} \bibinfo{author}{\bibfnamefont{V.}~\bibnamefont{Sachan}},
  \bibinfo{journal}{Phys. Rev. B} \textbf{\bibinfo{volume}{54}},
  \bibinfo{pages}{12318} (\bibinfo{year}{1996}).

\bibitem[{\citenamefont{Kajimoto et~al.}()\citenamefont{Kajimoto, Ishizaka,
  Yoshizawa, and Tokura}}]{kaji02}
\bibinfo{author}{\bibfnamefont{R.}~\bibnamefont{Kajimoto}},
  \bibinfo{author}{\bibfnamefont{K.}~\bibnamefont{Ishizaka}},
  \bibinfo{author}{\bibfnamefont{H.}~\bibnamefont{Yoshizawa}},
  \bibnamefont{and} \bibinfo{author}{\bibfnamefont{Y.}~\bibnamefont{Tokura}},
  \bibinfo{journal}{Phys. Rev. B} \textbf{\bibinfo{volume}{67}},
  \bibinfo{pages}{014511} (\bibinfo{year}{2003}).

\bibitem[{\citenamefont{Homes et~al.}(1993{\natexlab{b}})\citenamefont{Homes,
  Timusk, Liang, Bonn, and Hardy}}]{home93b}
\bibinfo{author}{\bibfnamefont{C.~C.} \bibnamefont{Homes}},
  \bibinfo{author}{\bibfnamefont{T.}~\bibnamefont{Timusk}},
  \bibinfo{author}{\bibfnamefont{R.}~\bibnamefont{Liang}},
  \bibinfo{author}{\bibfnamefont{D.~A.} \bibnamefont{Bonn}}, \bibnamefont{and}
  \bibinfo{author}{\bibfnamefont{W.~N.} \bibnamefont{Hardy}},
  \bibinfo{journal}{Phys. Rev. Lett.} \textbf{\bibinfo{volume}{71}},
  \bibinfo{pages}{1645} (\bibinfo{year}{1993}{\natexlab{b}}).

\bibitem[{\citenamefont{Zaanen et~al.}(2001)\citenamefont{Zaanen, Osman, Kruis,
  Nussinov, and {Tworzyd\l o}}}]{zaan01}
\bibinfo{author}{\bibfnamefont{J.}~\bibnamefont{Zaanen}},
  \bibinfo{author}{\bibfnamefont{O.~Y.} \bibnamefont{Osman}},
  \bibinfo{author}{\bibfnamefont{H.~V.} \bibnamefont{Kruis}},
  \bibinfo{author}{\bibfnamefont{Z.}~\bibnamefont{Nussinov}}, \bibnamefont{and}
  \bibinfo{author}{\bibfnamefont{J.}~\bibnamefont{{Tworzyd\l o}}},
  \bibinfo{journal}{Phil. Mag. B} \textbf{\bibinfo{volume}{81}},
  \bibinfo{pages}{1485} (\bibinfo{year}{2001}).

\bibitem[{\citenamefont{Chernyshev et~al.}(2000)\citenamefont{Chernyshev,
  {Castro Neto}, and Bishop}}]{cher00}
\bibinfo{author}{\bibfnamefont{A.~L.} \bibnamefont{Chernyshev}},
  \bibinfo{author}{\bibfnamefont{A.~H.} \bibnamefont{{Castro Neto}}},
  \bibnamefont{and} \bibinfo{author}{\bibfnamefont{A.~R.}
  \bibnamefont{Bishop}}, \bibinfo{journal}{Phys. Rev. Lett.}
  \textbf{\bibinfo{volume}{84}}, \bibinfo{pages}{4922} (\bibinfo{year}{2000}).

\bibitem[{\citenamefont{Romero et~al.}(1992)\citenamefont{Romero, Porter,
  Tanner, Forro, Mandrus, Mihaly, Carr, and Williams}}]{rome92}
\bibinfo{author}{\bibfnamefont{D.~B.} \bibnamefont{Romero}},
  \bibinfo{author}{\bibfnamefont{C.~D.} \bibnamefont{Porter}},
  \bibinfo{author}{\bibfnamefont{D.~B.} \bibnamefont{Tanner}},
  \bibinfo{author}{\bibfnamefont{L.}~\bibnamefont{Forro}},
  \bibinfo{author}{\bibfnamefont{D.}~\bibnamefont{Mandrus}},
  \bibinfo{author}{\bibfnamefont{L.}~\bibnamefont{Mihaly}},
  \bibinfo{author}{\bibfnamefont{G.~L.} \bibnamefont{Carr}}, \bibnamefont{and}
  \bibinfo{author}{\bibfnamefont{G.~P.} \bibnamefont{Williams}},
  \bibinfo{journal}{Phys. Rev. Lett.} \textbf{\bibinfo{volume}{68}},
  \bibinfo{pages}{1590} (\bibinfo{year}{1992}).

\bibitem[{\citenamefont{Lupi et~al.}(2000)\citenamefont{Lupi, Calvani, Capizzi,
  and Roy}}]{lupi00}
\bibinfo{author}{\bibfnamefont{S.}~\bibnamefont{Lupi}},
  \bibinfo{author}{\bibfnamefont{P.}~\bibnamefont{Calvani}},
  \bibinfo{author}{\bibfnamefont{M.}~\bibnamefont{Capizzi}}, \bibnamefont{and}
  \bibinfo{author}{\bibfnamefont{P.}~\bibnamefont{Roy}},
  \bibinfo{journal}{Phys. Rev. B} \textbf{\bibinfo{volume}{62}},
  \bibinfo{pages}{12418} (\bibinfo{year}{2000}).

\bibitem[{\citenamefont{Ando et~al.}(2001)\citenamefont{Ando, Lavrov, Komiya,
  Segawa, and Sun}}]{ando01}
\bibinfo{author}{\bibfnamefont{Y.}~\bibnamefont{Ando}},
  \bibinfo{author}{\bibfnamefont{A.~N.} \bibnamefont{Lavrov}},
  \bibinfo{author}{\bibfnamefont{S.}~\bibnamefont{Komiya}},
  \bibinfo{author}{\bibfnamefont{K.}~\bibnamefont{Segawa}}, \bibnamefont{and}
  \bibinfo{author}{\bibfnamefont{X.~F.} \bibnamefont{Sun}},
  \bibinfo{journal}{Phys. Rev. Lett.} \textbf{\bibinfo{volume}{87}},
  \bibinfo{pages}{017001} (\bibinfo{year}{2001}).

\end{thebibliography}

\end{document}